# Simulating Using Deep Learning The World Trade Forecasting of Export-Import Exchange Rate Convergence Factor During COVID-19

Effat Ara Easmin Lucky [1], Md. Mahadi Hasan Sany [1], Mumenunnesa Keya[1], Md. Moshiur Rahaman [1], Umme Habiba Happy[1], Sharun Akter Khushbu [1], Md. Arid Hasan [1]

[1] Department of Computer Science and Engineering, Daffodil International University, Dhaka 1209, Bangladesh

{effat15-10793, mahadi15-11173, mumenunnessa15-10100, moshiur15-11112, umme15-10916, sharun.cse, arid.cse0325.c}@diu.edu.bd

**Abstract.** By trade we usually mean the exchange of goods between states and countries. International trade acts as a barometer of the economic prosperity index and every country is overly dependent on resources, so international trade is essential. Trade is significant to the global health crisis, saving lives and livelihoods. By collecting the dataset called "Effects of COVID-19 on trade" from the state website NZ Tatauranga Aotearoa, we have developed a sustainable prediction process on the effects of COVID-19 in world trade using a deep learning model. In the research, we have given a 180-day trade forecast where the ups and downs of daily imports and exports have been accurately predicted in the Covid-19 period. In order to fulfill this prediction, we have taken data from 1st January 2015 to 30th May 2021 for all countries, all commodities, and all transport systems and have recovered what the world trade situation will be in the next 180 days during the Covid-19 period. The deep learning method has received equal attention from both investors and researchers in the field of in-depth observation. This study predicts global trade using the Long-Short Term Memory. Time series analysis can be useful to see how a given asset, security, or economy changes over time. Time series analysis plays an important role in past analysis to get different predictions of the future and it can be observed that some factors affect a particular variable from period to period. Through the time series it is possible to observe how various economic changes or trade effects change over time. By reviewing these changes, one can be aware of the steps to be taken in the future and a country can be more careful in terms of imports and exports accordingly. From our time series analysis, it can be said that the LSTM model has given a very gracious thought of the future world import and export situation in terms of trade.

# 1      Introduction

The world has become a global village and trade plays the main role between the countries as an interaction channel. Trade is a process of exchanging products and services between states or countries and moreover, it can see in between different companies, or any other substances. Trade plays an awfully important part in setting up a higher global economy. We can talk about China's trade as an example of world trade's importance. By addressing "World Factory" China is known in numerous nation's views to the world currently, and the trademark "Made in China" shows up in nearly every corner of the world [2]. Theories about worldwide trade expanding total efficiency at the nation's level are about as ancient as economics [3]. By the word 'world trade' or 'foreign trade', means that it is a process in which required goods or services are received and domestic products are sold in the global market [4]. Import and export are the two most important words in world trade, where import means purchasing foreign products or services and bringing them to one's home country and export means selling of products from home country to a foreign country. The relationship between foreign trade and economic growth is still a topic of debate from ancient times till now [5]. As a major factor of openness, worldwide trade has made a progressively significant contribution to economic growth [6]. This relationship (between foreign trade and economic growth) is based on two approaches- one is growth theory based on import and another one is growth theory based on export [7]. To capture the perfect view of the relationship between foreign trade and economic growth, time series analysis is always the best way [8]. The time series analysis method is utilized to analyze and understand past data and predict the future, generally, time-series data sets are annual data of different periods. When a set of observations is organized chronologically, then it is called a time series, time series analysis forms one of the most significant tools of the economist [9]. Long-term time series have various tricks or strategies like recursive, direct, and DirRec which are examined and compared [10]. In our paper, we collect the world trade data for the period January 01, 2015-May 30, 2021. The main purpose of the investigation is to compare the import and export data for all countries for the period January 1st, 2015-May 30, 2021, and based on the result the system predicts the future possibility of world trade. That's why we apply the time series model LSTM to analyze the data and prepare future predictions for the next 180 days of 2021.

# 2      Literature Review

Trade is an essential financial idea which counts the buying and selling of goods or projects and services, through paying payment by a buyer to a seller. From different papers, in this part, here reviewed the same categorical purposes. From the study, the findings are- a lot of variation in research with trade, many of them include here like, Avi et al. [11] investigate the global components of the economics of AI. In the discussion, trade theory which analyses the roles of scales, rivalry, and information creation, and information diffusion essential to near advantage comes up. Reuven et al. [12] describes a question in the work, the question is "Does a currency union affect trade?".

The description of it's given with time series and assesses the impact of currency unions on trade abusing time series variety and the data amount of trade was huge between 1970 and 1990. Both time series and cross-sectional variety of currency union rates had been utilized. Shahbaz et al. [13] explored how much vitality request is influenced by urbanization, financial improvement, and trade openness in essentialness request work within the case of Malaysia. They collected data from world development indicators and the data period was between 1971 to 2011, and by using the Interpolation method they convert annual data series into quarterly frequency. Jungho et al. [14] analyze the dynamic relationships between energy consumption, income growth, trade, and CO2 outflows for G-20 economies in a system of CVAR (cointegrated vector autoregression). The result shows that economic growth and trade have an inclinable impact on natural quality for the created G-20 portion countries. The data was collected from World Development Indicators (WDI). Shuanglin et al. [15] analyzed the trade and economic growth relationship and the data was China's national data from 1952 to 1997 and the output of their work was relatively positive with the growth rate of trade and growth rate of per capita GDP. Utku et al. [16] analyzed the need for imports and exports for Turkey concerning the EU. They applied time series analysis for the period 1963 to 2002 and used annual data. Their experimental results recommend that the Turkish import demand from the EU is obliged by its external obligation. Future prediction of trade is essential to know because it helps us to handle any circumstances. Rana et al. [17] examined the pattern of trade and estimated the future trend of Pakistan by using time series analysis for the period 1960-2010. The outcomes show that the export structure has changed essentially since the 1960s by a keen drop in the share of essential products and increment of industrial products. Exploring relationships between trade, FDI, and growth rate of per capita GDP is very important to the analysis. Mohammed et al. [18] investigated the relationship of the trade, foreign direct investment, and the growth rate of per capita GDP of Bangladesh from 1973 to 2014 with the help of time series analysis. The data was collected from the World Bank online database, their VECM model (Vector Error Correction Model) shows all of the variables of their paper have a long-term relationship. For the validity check of the VECM model, they applied several post estimation diagnostic tests and as result, they found that the residuals of the regressions have typical conveyance and don't appear to have any auto-correlation. Melina et al. [19] also applied time series analysis for the period 1960-2002 and analyzed the relationship of trade, foreign direct investment, and economic growth of Greece and as a result, they found a long-run balanced relationship between the mentioned variables. They collected annual data from the International Monetary Fund for the period 1960-2002 and applied the VAR (Vector Auto Regressive model) for determining the relationship between trade, foreign direct investment, and economic growth. Moreover, for the period of 1960-2005 in Turkey, the relationship between energy consumption, CO2 emissions, income, and foreign trade was analyzed by Halicioglu et al. [20]. They collected data from IMF (International Financial Statistics), and World Bank (WB) which is World development indicators, and Annual Statistics of Turkish Statistical Institute (TSI), and applied different models including the ARDL model. The experimental outcomes of their paper proposed that in Turkey, income is the most essential variable for the explanation of carbon emissions which is taken after by energy consumption and foreign trade.

## 3    Methodology

The workflow of our work is given below-

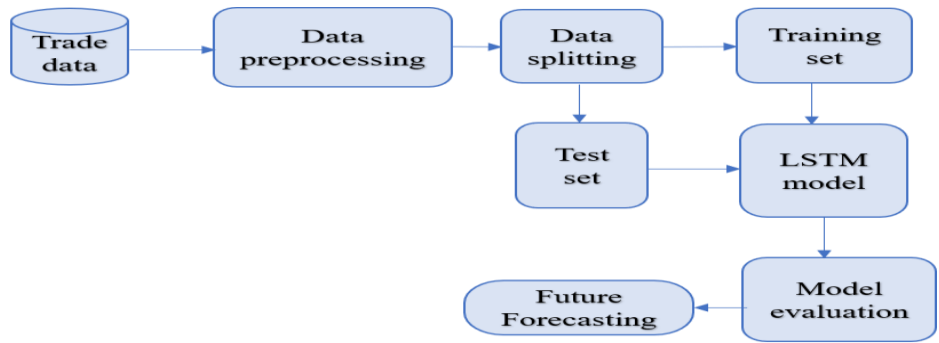

**Figure-1**: FlowChart of World Trade Time Series Predictor

### 3.1    Data Collection

Data collection is a process of gathering or collecting data and then estimating and analyzing them for a proper overview of research utilized standard endorsed procedure. Data collection is the very first and most important part of any research work. There are many different processing of collecting data based on the research work. We can collect data from different social networking media and websites [21]. And thus, online data collection is becoming a progressively well-known research strategy [22]. For the work, we collect trade data from the States NZ Tatauranga Aotearoa website. There are import and export data from January 1st, 2015-May 30, 2021 for all countries, all commodities, and all transport systems. Then we analyze these data and predict the future for the next 180 days. The daily basis global import & export values are shown below in figure-2.

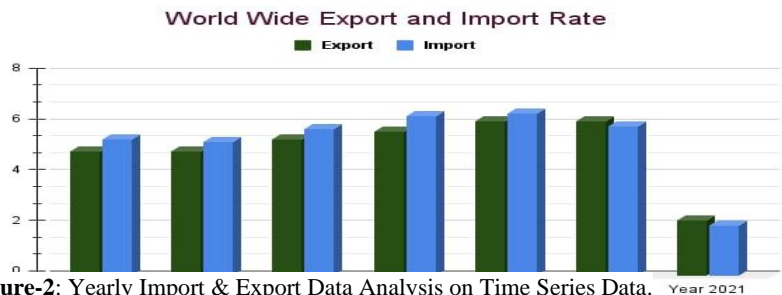

**Figure-2**: Yearly Import & Export Data Analysis on Time Series Data.

### 3.2 Long-Short Term Memory

Long Short-Term Memory is known as a type of artificial recurrent neural network and it is utilized in the deep learning field for providing long-term association. LSTM works better than traditional recurrent neural networks (RNN) [23]. It was influenced by the analysis of error flow in the existing RNN. LSTM has hidden layers and in hidden layers, there is a set of blocks which are known as memory blocks, these blocks have differentiable memory chips in advanced computers [24]. These memory chips have one or more repetitively involved memory cells and also have three multiplicative units including input, output, and input gates. And these units provide continuous analogs of reading, write and reset operations for the cells. Net is used for interacting with the cells through the gates. We use adam optimizer because it attaches the most excellent properties AdaGrad and RMSProp algorithms to supply an optimization algorithm that can manage inadequate slopes on riotous issues. The adam optimizer is much faster than other optimization methods in neural networks [25]. We also use Mean Square Error for minimizing loss.

**Table-1**: Table (a) & (b) Represent the LSTM Model Summary

| Layer(type) | Output shape | Param |
|---|---|---|
| lstm_6 (LSTM) | (1, 1, 4) | 96 |
| lstm_7 (LSTM) | (1, !) | 24 |
| dense_3(Dense) | (1, !) | 2 |

Table-(a)

| Total params | Trainable Params | Non-trainable params |
|---|---|---|
| 122 | 122 | 0 |

Table-(b)

### 3.3 LSTM Model Estimation with the Parameters

The parameters of the model are set in the LSTM model very easily with the appropriate input in real time transmission. We work with 60% and 30% data for training and testing and where batch size and epoch are 1 and 100 respectively.

**Table-2**: LSTM Transcendence Order

| LSTM Transcendence Order | | | |
|---|---|---|---|
| Training Size | Testing Size | Batch Size | Epochs |
| 70% | 30% | 1 | 100 |

### 3.4 LSTM Model Analysis

For a strong representation of future predictions, a comparative study is very essential. From the dataset of world trade, significant with LSTM where the Long and Short-Term groupings and it can be extricated the procedure. We trained the export-import trade data from January 1st, 2015 to May 30, 2021, per day. World trade study prediction is trained by LSTM which gives the actual prediction [26].

Table- 3: Model loss on Gobal Exchange Rate.

|  | EXPORT | | IMPORT | |
|---|---|---|---|---|
|  | **Highest loss** | **Lowest loss** | **Highest loss** | **Lowest loss** |
| **ALL OVER WORLD** | 0.0507 | 0.0378 | 0.2652 | 0.0081 |

## 4 RESULTS AND DISCUSSION

Figure-3 is visually representing the training testing values globally for import. The blue, yellow, green, and red waves give the train (true & predicted value), test (true & predicted value) consequently. We can see that the training true value and testing predict values are relatively close. The training testing values globally for Imports start from 2015 and end in 2021.

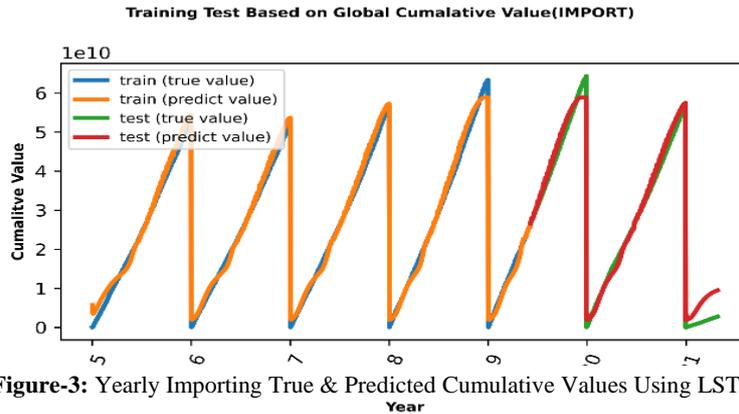

**Figure-3:** Yearly Importing True & Predicted Cumulative Values Using LSTM.

Figure-4 has plotted the training testing values globally for export. In this graph, the blue wave indicates the train (true value), the yellow wave indicates train (predict value), the green wave indicates test (true value), and the red wave indicates test (predict value). That train (true value) is relatively close to the test (true value), and the train (predict value) is relatively close to the test (predict value). The training testing values globally for exports start from 2015 and end in 2021.

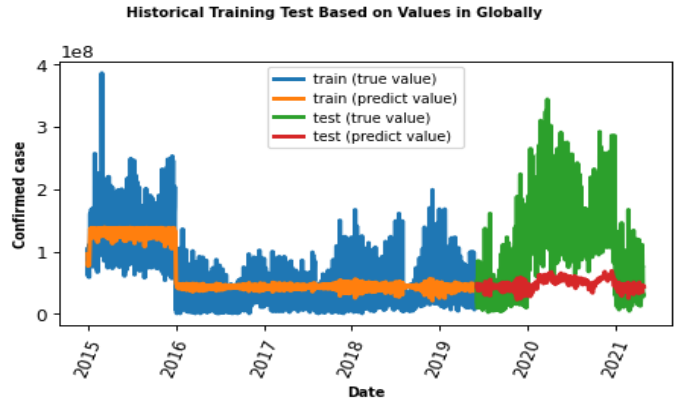

**Figure-4:** Yearly Exporting True & Predicted Cumulative Values Using LSTM

In figure-5, we can see the future prediction on per day Cumulative values of global import. In this graph, it has seen that the value 2.56 is constant with the date. The beginning date of the prediction is 2021-05 and the ending date is 2021-11.

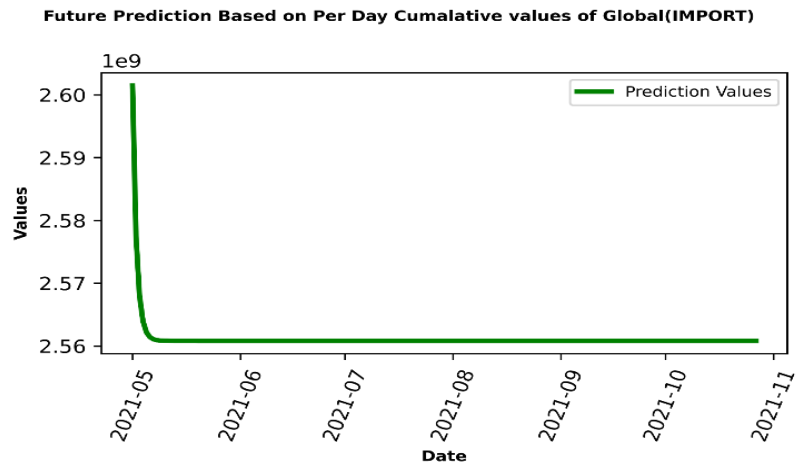

**Figure-5**: Future Trajectory Based on Per Day Import Cumulative Values.

Figure-6 visualizes the future prediction on per day Cumulative values of global export. Future prediction for each day is shown in the given graph. It is also seen that the value 4.94 is constant for a long time. The beginning date of the prediction is 2021-05 and the ending date is 2021-11.

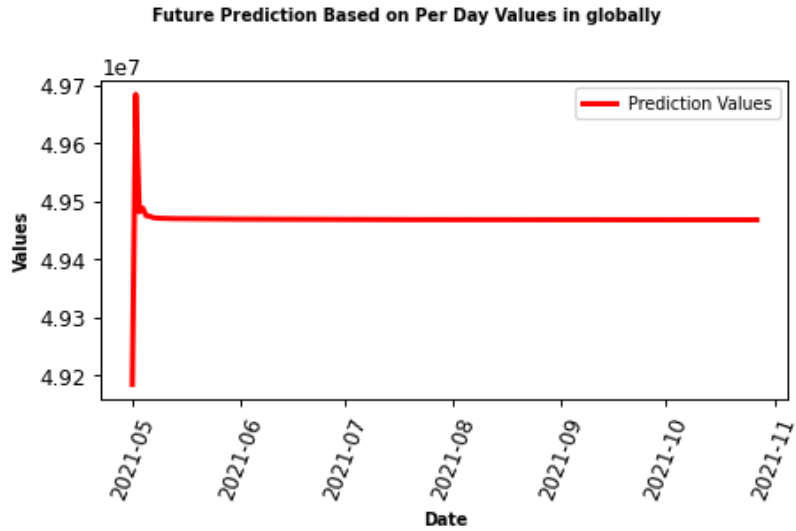

**Figure-6**: Future Trajectory Based on Per Day Export Cumulative Values.

Figure-7 has plotted the future prediction of global imports for the next 180 days. It shows that the result will increase constantly in the next 180 days (from June 2021 to November 2021). By comparing the previous data and our result, we can say that our model estimation is so much true.

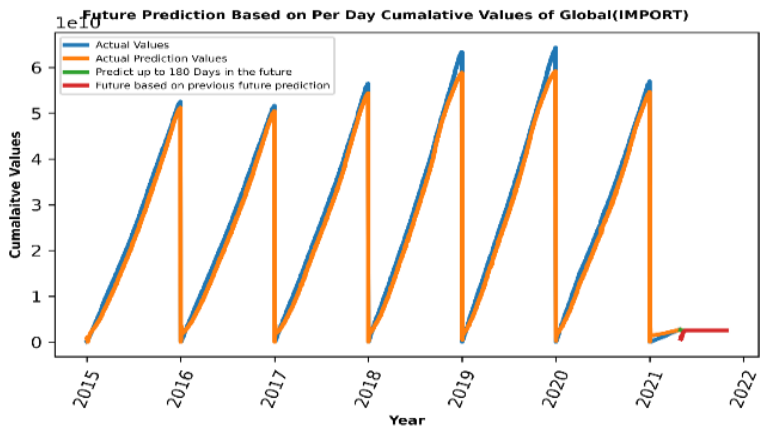

**Figure-7**: Future Prediction Based on Per Day Import Cumulative Values for Next Six Months UsingLSTM.

Figure-8 has plotted the future prediction of global exports for the next 180 days. It shows that the result will increase constantly in the next 180 days. We can say that our model estimation is so true after comparing the previous data and our result.

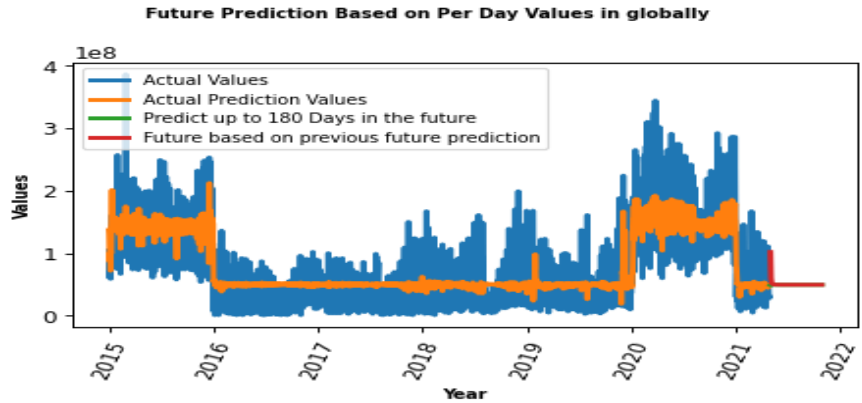

**Figure-8:** Future Prediction Based on Per Day Export Cumulative Values for Next Six Months Using LSTM

We wanted to predict future for world trade. And we are successful on our work. Our main focus/choice is to predict future by using time series analysis and we have done it smoothly by applying LSTM model. By comparing our results with more relatable works result [16-17], we can say that in both result, future prediction have done with LSTM and time series analysis.

## 5 Conclusion and future work

This research provides an idea of what the future holds for imports and exports in world trade during the Corona period with the LSTM model. The dataset, titled "Effects of COVID-19 on trade", was collected from the state website NZ Tatauranga Aotearoa to provide time-series analysis and is the original source of the data. The time series analysis has been studied based on export and import data from 2015, January 1 to 2021, May 30, for all countries, all commodities, and all transport systems. In the methodology, we have tried to predict what might happen in the next 180 days in world trade through various graphical visualizations which have used cumulative value in global trade in terms of imports and exports. The 180-day forecast shows that in the Covid-19 period the cumulative value of imports and exports is likely to remain unchanged from May 30, 2021 to November 2021.

We have some limitations in this work as we have only worked and predicted cumulative value but what the global economic situation might be in terms of trade was not discussed during this coronal period or no graphical representation was shown. In future, we'll try to create comparative study between the time series analysis of LSTM & ARIMA, SRIMA model.